\documentclass{article}
\usepackage{amsmath, amssymb, amsfonts}
\title{Hyperbolic motion generated by inversion} 
\author{Hristu Culetu\\ Ovidius University, Dept.of Physics, B-dul Mamaia 124, \\ 8700 Constanta, Romania \\ email : hculetu@yahoo.com}
\begin{document}
\numberwithin{equation}{section}
\pagenumbering{arabic}
\maketitle
\begin{abstract}
An inversion transformation applied to an inertial observer is used to generate a nonstatic conformally flat geometry in spherical coordinates. A static observer in the new geometry is uniformly accelerating with respect to the inertial one and vice versa, but its acceleration $g$  undergoes the transformation $g \rightarrow 1/b^{2}g$, where $b$ is a constant. A nongeodesic congruence of a static observer has a scalar expansion which grows linearly with time but the acceleration is proportional to $r$, as for the classical rotation.
\end{abstract}
\newcommand{\fv}{\boldsymbol{f}}
\newcommand{\tv}{\boldsymbol{t}}
\newcommand{\gv}{\boldsymbol{g}}
\newcommand{\OV}{\boldsymbol{O}}
\newcommand{\wv}{\boldsymbol{w}}
\newcommand{\WV}{\boldsymbol{W}}
\newcommand{\NV}{\boldsymbol{N}}
\newcommand{\hv}{\boldsymbol{h}}
\newcommand{\yv}{\boldsymbol{y}}
\newcommand{\RE}{\textrm{Re}}
\newcommand{\IM}{\textrm{Im}}
\newcommand{\rot}{\textrm{rot}}
\newcommand{\dv}{\boldsymbol{d}}
\newcommand{\grad}{\textrm{grad}}
\newcommand{\Tr}{\textrm{Tr}}
\newcommand{\ua}{\uparrow}
\newcommand{\da}{\downarrow}
\newcommand{\ct}{\textrm{const}}
\newcommand{\xv}{\boldsymbol{x}}
\newcommand{\mv}{\boldsymbol{m}}
\newcommand{\rv}{\boldsymbol{r}}
\newcommand{\kv}{\boldsymbol{k}}
\newcommand{\VE}{\boldsymbol{V}}
\newcommand{\sv}{\boldsymbol{s}}
\newcommand{\RV}{\boldsymbol{R}}
\newcommand{\pv}{\boldsymbol{p}}
\newcommand{\PV}{\boldsymbol{P}}
\newcommand{\EV}{\boldsymbol{E}}
\newcommand{\DV}{\boldsymbol{D}}
\newcommand{\BV}{\boldsymbol{B}}
\newcommand{\HV}{\boldsymbol{H}}
\newcommand{\MV}{\boldsymbol{M}}
\newcommand{\be}{\begin{equation}}
\newcommand{\ee}{\end{equation}}
\newcommand{\ba}{\begin{eqnarray}}
\newcommand{\ea}{\end{eqnarray}}
\newcommand{\bq}{\begin{eqnarray*}}
\newcommand{\eq}{\end{eqnarray*}}
\newcommand{\pa}{\partial}
\newcommand{\f}{\frac}
\newcommand{\FV}{\boldsymbol{F}}
\newcommand{\ve}{\boldsymbol{v}}
\newcommand{\AV}{\boldsymbol{A}}
\newcommand{\jv}{\boldsymbol{j}}
\newcommand{\LV}{\boldsymbol{L}}
\newcommand{\SV}{\boldsymbol{S}}
\newcommand{\av}{\boldsymbol{a}}
\newcommand{\qv}{\boldsymbol{q}}
\newcommand{\QV}{\boldsymbol{Q}}
\newcommand{\ev}{\boldsymbol{e}}
\newcommand{\uv}{\boldsymbol{u}}
\newcommand{\KV}{\boldsymbol{K}}
\newcommand{\ro}{\boldsymbol{\rho}}
\newcommand{\si}{\boldsymbol{\sigma}}
\newcommand{\thv}{\boldsymbol{\theta}}
\newcommand{\bv}{\boldsymbol{b}}
\newcommand{\JV}{\boldsymbol{J}}
\newcommand{\nv}{\boldsymbol{n}}
\newcommand{\lv}{\boldsymbol{l}}
\newcommand{\om}{\boldsymbol{\omega}}
\newcommand{\Om}{\boldsymbol{\Omega}}
\newcommand{\Piv}{\boldsymbol{\Pi}}
\newcommand{\UV}{\boldsymbol{U}}
\newcommand{\iv}{\boldsymbol{i}}
\newcommand{\nuv}{\boldsymbol{\nu}}
\newcommand{\muv}{\boldsymbol{\mu}}
\newcommand{\lm}{\boldsymbol{\lambda}}
\newcommand{\Lm}{\boldsymbol{\Lambda}}
\newcommand{\opsi}{\overline{\psi}}
\renewcommand{\tan}{\textrm{tg}}
\renewcommand{\cot}{\textrm{ctg}}
\renewcommand{\sinh}{\textrm{sh}}
\renewcommand{\cosh}{\textrm{ch}}
\renewcommand{\tanh}{\textrm{th}}
\renewcommand{\coth}{\textrm{cth}}

 The hyperbolic motion is especially related to the long standing problem of the electromagnetic radiation emitted by a uniformly accelerated charge. Eminent scientists such as M. Born, W. Pauli, M. von Laue, etc. have given important contributions concerning the existence of the radiation emitted by a uniformly accelerated charge and its relationship with the Equivalence Principle. 
 
 Fulton and Rohrlich \cite{FR} have shown that the hyperbolic motion is related to the special conformal transformation, as a part of the 15 parameters conformal group, including the 10 parameters of the Lorentz group. It contains also an extra dilatational parameter and another four characterizing the special conformal transformation (see also \cite{EH, DB, CC})
 \begin{equation}
 x^{'b} = \frac{x^{b} - a^{b} x_{c}x^{c}}{1 - 2a_{b} x^{b} + a^{2} x_{b} x^{b}}
\label{1}
\end{equation}
where $x^{c}x_{c} = \eta_{ab} x^{a}x^{b}$ is the Minkowski interval, $\eta_{ab} = diag(-1, 1, 1, 1)$ and $a^{b}$ is a constant vector with dimension of acceleration \cite{HC}. The latin indices run from 0 to 3 and we take $G = c = 1$. The coordinate transformation (1) becomes the Newtonian acceleration transformation when $ax << 1, x_{c}x^{c} << 1/a$, which leads to $x' = x - at^{2}$ (for unidimensional motion), with $a^{b} = (0, a, 0, 0)$. It is composed by the following set (inversion - translation - inversion) 
 \begin{equation}
 x^{b} \rightarrow \frac{x^{b}}{x^{a}x_{a}} \rightarrow \frac{x^{b}}{x^{a}x_{a}} - a^{b} \rightarrow x^{'b}
\label{2}
\end{equation}
 We suggest in this letter to arrive at an uniformly accelerated motion using a different approach, in spherically-symmetric coordinates. Liu et al \cite{LCN} studied the problem of the quasi-local energy and its dependence on the choice of the reference system. They introduced the Minkowski metric in the form
 \begin{equation}
 ds^{2} = -dT^{2} + dR^{2} + R^{2} (d\theta^{2} + sin^{2}\theta d\phi^{2})
\label{3}
\end{equation}
 and performed the  transformation
  \begin{equation}
  T = T(r, t), ~~~~R = R(r, t),
\label{4}
\end{equation}
leaving the angular coordinates unchanged. One obtains
\begin{equation}
 ds^{2} = - (\dot{T}^{2} - \dot{R}^{2}) dt^{2} + 2 (\dot{R} R' - \dot{T} T') dt dr + (R^{'2} - T^{'2}) dr^{2} + R^{2} d\Omega^{2}, 
\label{5}
\end{equation}
with $d \Omega^{2} = d\theta^{2} + sin^{2}\theta d\phi^{2},~\dot{T} = \partial T/\partial t,~~T' = \partial T/\partial r$ and similar relations for $R$. 

Our goal now is to choose the functions $R$ and $T$ from (4) in order that the spacetime (5) to become conformally-flat. Therefore, we should have
 \begin{equation}
 \dot{T}^{2} - \dot{R}^{2} = R^{'2} - T^{'2} = \frac{R^{2}}{r^{2}} ,~~~~\dot{R} R' - \dot{T} T' = 0.
\label{6}
\end{equation}
Using common algebra and suitable initial conditions, we find that
  \begin{equation}
   T(r, t) = \frac{b^{2} t}{r^{2} - t^{2}}   , ~~~~ R(r, t) = \frac{b^{2} r}{r^{2} - t^{2}},
\label{7}
\end{equation}
where $r > t$ and the constant $b$ (with dimension of length) was introduced for dimensional reasons. It is worth noting that the functions $r(T, R)$ and $t(T, R)$ have the same form as $R(t, r)$ and, respectively, $T(t, r)$ from eqs. (7). The metric (5) yields now 
 \begin{equation}
 ds^{2} = \frac{b^{4}}{(r^{2} - t^{2})^{2}} (-dt^{2} + dr^{2} + r^{2} d\Omega^{2})
\label{8}
\end{equation}
which is a spherically symmetric conformally flat spacetime. The transformations (7) look like inversions (see (2)). Therefore, the relations between the corresponding Minkowski intervals is given by 
  \begin{equation}
 r^{2} - t^{2} =  \frac{b^{4}}{R^{2} - T^{2}}, ~~~~R > T.
\label{9}
\end{equation}
That means the constant $b$ is the geometrical mean of the two Minkowski intervals. In addition, a hyperbolic motion $R^{2} - T^{2} = 1/g^{2}$ in Minkowski space generates a similar  hyperbolic motion $r^{2} - t^{2} = (b^{2}g)^{2}$, with the new constant acceleration $1/b^{2}g$. We note also that the conformal factor from (8) is Lorentz-invariant and the spacetime possesses a horizon when $r^{2} - t^{2}$ tends to infinity or when $b^{2}/(r^{2} - t^{2})$ tends to zero. That can be seen from the relation (9) where $r^{2} - t^{2} \rightarrow \infty$ leads to $R^{2} - T^{2} \rightarrow 0$, namely the light cone in Minkowski coordinates. As is well known, it corresponds to the Rindler horizon in Rindler's coordinates. 

Let us find now the equation of motion of a static ($r = r_{0}$) particle from the point of view of an inertial (Minkowskian) observer. Taking $r = r_{0} = const.$ in (7) and eliminating the time $t$, one finds
  \begin{equation}
  R - \frac{b^{2}}{2r_{0}} = \sqrt{T^{2} + (\frac{b^{2}}{2r_{0}})^{2}}
\label{10}
\end{equation}
 where the plus sign has been chosen in front of the square root to get $R > 0$. The curves (10) are hyperbolas, with constant acceleration $2r_{0}/b^{2}$. 
 
 We have here a set of expanding particles (going away from the origin of the $(R, T)$ (or $(r, t)$)  system, starting from the sphere $R = b^{2}/r_{0}$ at $T = 0$ and moving with an acceleration which depends on the value of its fixed position in the $(r, t)$ system. In other words, the spacetime (8) corresponds to uniformly accelerated observers. Put it differently, the inversion (7) takes a particle from rest and moves it hyperbolically. We could therefore consider that the geometry (8) in spherical coordinates plays a similar role with the Rindler geometry in rectangular coordinates. Unfortunately, (8) is nonstatic and we have no a timelike Killing vector to define a conserved energy.
 
 An inertial particle located at $R = R_{0}$ will move, of course, on a hyperbolic trajectory (because of the reciprocity of the relations (7)). We have, in the $(r, t)$ system
   \begin{equation}
  r - \frac{b^{2}}{2R_{0}} = \sqrt{t^{2} + (\frac{b^{2}}{2R_{0}})^{2}}
\label{11}
\end{equation}
where the constant acceleration is now $2R_{0}/b^{2}$. Until now the constant $b$ is arbitrary. It seems to be related to the scale of the acceleration (the smaller $b$ is, the larger the acceleration). 

Our next task is to study the kinematical parameters of a congruence of static observers in our spacetime (8). Let us choose the velocity field $u^{a} = ((r^{2} - t^{2})/b^{2},~ 0,~ 0,~ 0)$, with $u^{a} u_{a} = -1$. The scalar expansion of the congruence is 
   \begin{equation}
   \Theta \equiv \nabla_{a} u^{a} = \frac{1}{\sqrt{-g}} \frac{\partial}{\partial x^{a}} (\sqrt{-g} u^{a}) = \frac{6t}{b^{2}}
\label{12}
\end{equation}
where $g$ is here the determinant of the metric. For the time evolution of the expansion we have
   \begin{equation}
   \dot{\Theta} \equiv  u^{a} \nabla_{a}\Theta = \frac{6(r^{2} - t^{2})}{b^{4}},
\label{13}
\end{equation}
which is always positive.

The congruence $u^{a}$ is not geodesic. It can be seen by computing its acceleration, which has the components
   \begin{equation}
  a^{b} \equiv  u^{a} \nabla_{a} u^{b} = (0,~ \frac{-2r(r^{2} - t^{2})}{b^{4}},~ 0,~ 0),
\label{14}
\end{equation}
In addition, eq. (14) yields
   \begin{equation}
  \nabla_{b}~ a^{b} = \frac{6(r^{2} + t^{2})}{b^{4}}, ~~~~a \equiv \sqrt{a^{b} a_{b}} = 2r/b^{2}. 
\label{15}
\end{equation}
One means the acceleration $a$ is proportional to the distance from the origin $r = 0$, as in the case of the classical rotation. Therefore, the constant $b$ could be related to the frequency $\omega$ of a rotation with $\omega = \sqrt{2}/b$. Because of the spherical symmetry, the shear tensor $\sigma_{ab}$ of the congruence vanishes. The same is valid for the vorticity tensor $\omega_{ab}$. We could now check that the Raychaudhuri equation 
   \begin{equation}
  \dot{\Theta} + \frac{\Theta^{2}}{3} +\sigma^{2} - \omega^{2} - \nabla_{b} a^{b} = - R_{ab} u^{a} u^{b} 
\label{16}
\end{equation}
is satisfied. We have here $2\sigma^{2} = \sigma_{ab} \sigma^{ab},~2\omega^{2} = \omega_{ab} \omega^{ab}$ and $R_{ab} = 0$ because the metric (8) is flat.

 To summarize, we studied in this letter the hyperbolic motion in spherical coordinates and found a reciprocal relation between the equations of motion obtained with respect to the inertial and accelerated reference system generated by inversion. The relation between accelerations is reciprocal, too. For a nongeodesic congruence of static particles we obtained $\Theta \propto t$ and $\sqrt{a_{b} a^{b}} \propto r$.

\end{document}